\documentclass[journal=jpcl,manuscript=extarticle,layout=twocolumn]{achemso}
\setkeys{acs}{maxauthors=0}

\let\titlefont\undefined
\makeatletter
\let\l@addto@macro\relax
\makeatother
\usepackage[fontsize=10pt]{scrextend}

\usepackage{graphicx}%
\usepackage{bm}%
\usepackage{xcolor}
\usepackage{siunitx}
\usepackage{physics}
\usepackage{appendix}
\usepackage{nicefrac}

\usepackage{soul}

\usepackage{soul} %

\usepackage{amssymb}

\usepackage[symbol]{footmisc}

\usepackage{multicol}
\usepackage{ragged2e}
\usepackage{float}
\newfloat{textbox}{tbp}{lop}
\floatname{textbox}{Text Box}
\usepackage{afterpage}
\usepackage{dblfloatfix}
\usepackage{placeins}
\usepackage{cuted}

\usepackage{braket}
\usepackage{dcolumn} %
\newcolumntype{d}[1]{D{.}{.}{#1}} %

\everymath{\small} %
\everydisplay{\normalsize}    %

\title{Artificial Thermalization in Ring-Polymer Molecular Dynamics: The Breakdown of RPMD for Gas-Phase Reactions with Pre-Reactive Complexes and How to Fix It}

\author{Joseph E.\ Lawrence}
\email{joseph.lawrence@nyu.edu}
\affiliation{Department of Chemistry, New York University, New York, NY 10003, USA\\Simons Center for Computational Physical Chemistry, New York University, New York, NY 10003, USA}
\alsoaffiliation{Department of Chemistry and Applied Biosciences, ETH Zurich, 8093 Zurich, Switzerland}
\author{Jeremy~O.\ Richardson}
\email{jeremy.richardson@phys.chem.ethz.ch}

\affiliation{Department of Chemistry and Applied Biosciences, ETH Zurich, 8093 Zurich, Switzerland}

\date{\today}%

\begin{document}

\maketitle

\begin{strip}
    \centering
    \begin{minipage}{1.0\textwidth}
\begin{multicols}{2}
\includegraphics[width=1.0
\linewidth]{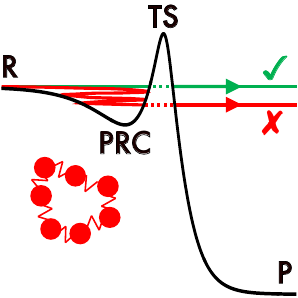}
\vspace{1cm}

\justifying
   \noindent{\bf Abstract:} %
   Ring-polymer molecular dynamics (RPMD) has become a popular method for describing chemical reactions due to its ability to simultaneously capture tunneling, zero-point energy, anharmonicity and recrossing. Here we highlight that despite its many successes, great care must be taken when applying RPMD to study gas-phase reactions at low pressure. We show that for bimolecular reactions that proceed via pre-reactive complexes, RPMD predicts spuriously large rates at low temperatures and pressures. Using the rigorous connection of RPMD and semiclassical instanton theory, we demonstrate that this breakdown can be understood in terms of an intrinsic problem with RPMD: artificial thermalization. In the present context, this opens up reactive channels below the reactant asymptote that should be energetically inaccessible, resulting in erroneously large rates. We discuss practical strategies to overcome this problem by combining the steepest-descent inverse Laplace transform with Bleistein's uniform approximation to calculate %
   the thermal rate given an appropriate lower energy bound.
\end{multicols} 
\vspace{-0.5cm}
\noindent\rule{\textwidth}{0.4pt}
        \vspace{-0.8cm}
    \end{minipage}
\end{strip}

 \noindent \textbf{Introduction:}  %
 Accurately modeling chemical reaction rates presents a number of challenges.
 Not only must one use a sufficiently high level of electronic-structure theory, but
 one must also accurately describe the dynamics on the resulting surface. For systems with a small number of degrees of freedom, exact nuclear wavefunction methods can be used.\cite{Skouteris2000ABC,Althorpe+Clary2003review,Wu2004HCH4,Hoppe2024Cl+CH4,Song2023Submerged_Barriers} However,  the exponential scaling of quantum mechanics means that, even for moderately sized gas-phase reactions, exact methods in full atomistic detail are prohibitively expensive. Unfortunately, classical dynamics is often insufficient, due to the importance of nuclear tunneling and zero-point energy.

Ring-polymer molecular dynamics (RPMD) is a powerful approach for simulating reaction rates in full atomistic detail while incorporating anharmonicity and nuclear quantum effects.\cite{RPMDcorrelation,RPMDrate,RPMDrefinedRate,Habershon2013RPMDreview} It is based on the isomorphism between the quantum statistical mechanics of a system in thermal equilibrium and the classical statistical mechanics of ring polymers, consisting of $n$ copies of the corresponding classical system connected by harmonic springs.\cite{Chandler+Wolynes1981,Parrinello1984Fcenter} RPMD is simply classical dynamics in the extended phase space of the ring polymer. Although only approximate, the dynamics satisfies a number of key properties in the $n\rightarrow\infty$ limit:\cite{RPMDcorrelation,Lawrence2020rates}  it conserves the exact quantum Boltzmann distribution, it obeys detailed balance, it is accurate to at least $\mathcal{O}(t^3)$ for correlation functions of operators that are functions of position,\cite{Braams2006RPMD,Jang2014RPMD,Rossi2014resonance} and it correctly reduces to classical dynamics in the high-temperature limit. These properties make RPMD well suited to the simulation of large condensed-phase systems.
Crucially, as well as being accurate at short time, RPMD is capable of predicting long-time dynamical properties such as rate constants and diffusion coefficients.\cite{RPMDrate,RPMDrefinedRate,RPMDprotonTransfer,RPMDdiffusion,Markland2008HandMuDiffusionIceWater,Habershon2009qtip4pf,Cendagorta2016clathrate}  
 The accuracy of RPMD for reaction rates has not only been demonstrated numerically,\cite{RPMDgasPhase,Suleimanov2011HCH4,Suleimanov2013RPMDrate,Suleimanov2013H+H2,DMuH,Suleimanov2016rate} but can also be rationalized in terms of its connection to instanton theory.\cite{Richardson2009RPInst}

 Instanton theory\cite{Miller1975semiclassical,Chapman1975rates,Coleman1977ImF,Callan1977ImF,Coleman1979UsesOfInstantons,Stone1977ImF,Affleck1981ImF,Richardson2016FirstPrinciples,Richardson2018InstReview,Lawrence2024crossover} is a rigorous semiclassical ($\hbar\to0$) approximation to the exact quantum rate.\cite{Miller1983rate} The instanton is the dominant tunneling path in a path-integral description of the reaction rate constant.\cite{Perspective} This path can  be interpreted as a periodic classical trajectory in imaginary time, equivalent to a real-time trajectory on the upturned potential.\cite{Miller1975semiclassical} Importantly, for the connection to RPMD, the instanton is also equivalent to the saddle point on the ring-polymer potential, i.e.~the optimum ring-polymer transition state.\cite{Richardson2009RPInst}  
 In fact, the harmonic approximation to the optimum ring-polymer transition-state theory differs from instanton theory only slightly in its treatment of the unstable mode.
 Furthermore, the resulting terms can be shown to be equivalent at the crossover temperature, i.e.~the temperature at which the instanton first appears.
 This explains why RPMD rate theory not only describes shallow tunneling but also typically gives an accurate description of deep tunneling.\cite{Richardson2009RPInst}  

Although originally developed with condensed-phase simulation in mind, RPMD has been successfully applied to many elementary gas-phase reactions.\cite{RPMDgasPhase,Suleimanov2011HCH4,Suleimanov2013RPMDrate,Suleimanov2013H+H2,DMuH,Suleimanov2016rate} 
Following these early successes, in the past few years, 
RPMD has  been applied to study  a range of gas-phase reactions exhibiting pre-reactive complexes\cite{delMazo2021OH+H2CO,Espinosa2020H_Cl_F_Ethane,Novikov2024PRC_RPMD_OH+HBr,Murakami2024PRC_RPMD_H2CO-+CH3Cl,Murakami2024PRC_OH+CH3OH_photodissociation_study,EspinosaGarcia2024PRC_RPMD_CN+Ethane,delMazoSevillano2023PRC_RPMD_(H2CO)2+OH,Hashimoto2023RPMDNH3H2,Murakami2022PRC_RPMD_H-+C2H2,Espinosa2022PRC_RPMD_OH+SiH4,Braunstein2022PRC_RPMD_D+H3+,Saito2021PRC_RPMD_CO+H3+_BranchingRatios,Cao2021PRC_RPMD_C+H2,Yang2021Roaming_RPMD_H+MgH,Chen2021PRC_RPMD_H+O3,Liu2020PRC_RPMD_OH+HO2,Song2020PRC_RPMD_OH+HO2,Bulut2019RPMDPreReactiveComplex,Naumkin2019PRC_RPMD_OH+Methanol_and_Formaldehyde,delMazo2019PRC_RPMD_OH+Methanol_and_Formaldehyde,Suleimanov2018RPMD_H3++H2,EspinosaGarcia2017PRC_RPMD_CN+CH4,Zuo2017PRC_RPMD_OH+HCl,Zuo2016PRC_RPMD_OH+HCl} (potential energy wells before the reaction barrier).
Such reactions are important in a number of contexts; one area in which they play a particularly interesting role is  
in the formation of organic molecules in the interstellar medium.\cite{Shannon2013tunnel,Zhang2022Faraday,Puzzarini2022InterStellarMediumHardInLab}
On the basis of classical transition-state theory, one would expect that at the low temperatures of deep space, the presence of even very small reaction barriers should lead to negligibly slow reactions.
However, the presence of a pre-reactive well enables the reactants to become trapped as a pre-reactive complex that can live long enough for the system to tunnel through the barrier to the products.\cite{Shannon2013tunnel}
Gas-phase reactions are typically described as being in the low-pressure limit if tunneling occurs faster than collisions with other molecules, or the high-pressure limit if the system is thermalised in the pre-reactive well.  In this paper, we are interested in the low-pressure limit.

The importance of tunneling, zero-point energy and anharmonicity make it seem, \emph{prima facie}, that these reactions would be an ideal application for RPMD\@. 
However, as we shall demonstrate in the following, the presence of the pre-reactive complex actually leads to a breakdown of RPMD at low temperatures and pressures. Some previous studies have already noticed issues empirically,\cite{delMazoSevillano2023PRC_RPMD_(H2CO)2+OH,delMazo2021OH+H2CO,Hashimoto2023RPMDNH3H2} and tentative explanations have been suggested\cite{delMazo2021OH+H2CO,delMazoSevillano2023PRC_RPMD_(H2CO)2+OH} based on the so-called ``spurious resonances'' exhibited in RPMD vibrational spectra.\cite{Witt2009spectroscopy,Rossi2014resonance}
In this letter, we use a simple model system to demonstrate the problem. We then analyze the breakdown using the connection of RPMD to instanton theory. Rather than ``spurious resonances'', we find that the cause of the breakdown is artificial thermalization, which results in the contamination of the rate by low-energy states in the pre-reactive well below the reactant asymptote. This analysis will lead to a simple suggestion for how the breakdown can be overcome.
\begin{figure}[t]
    \centering
    \vspace{-0.5cm}\includegraphics[width=1.0\linewidth]{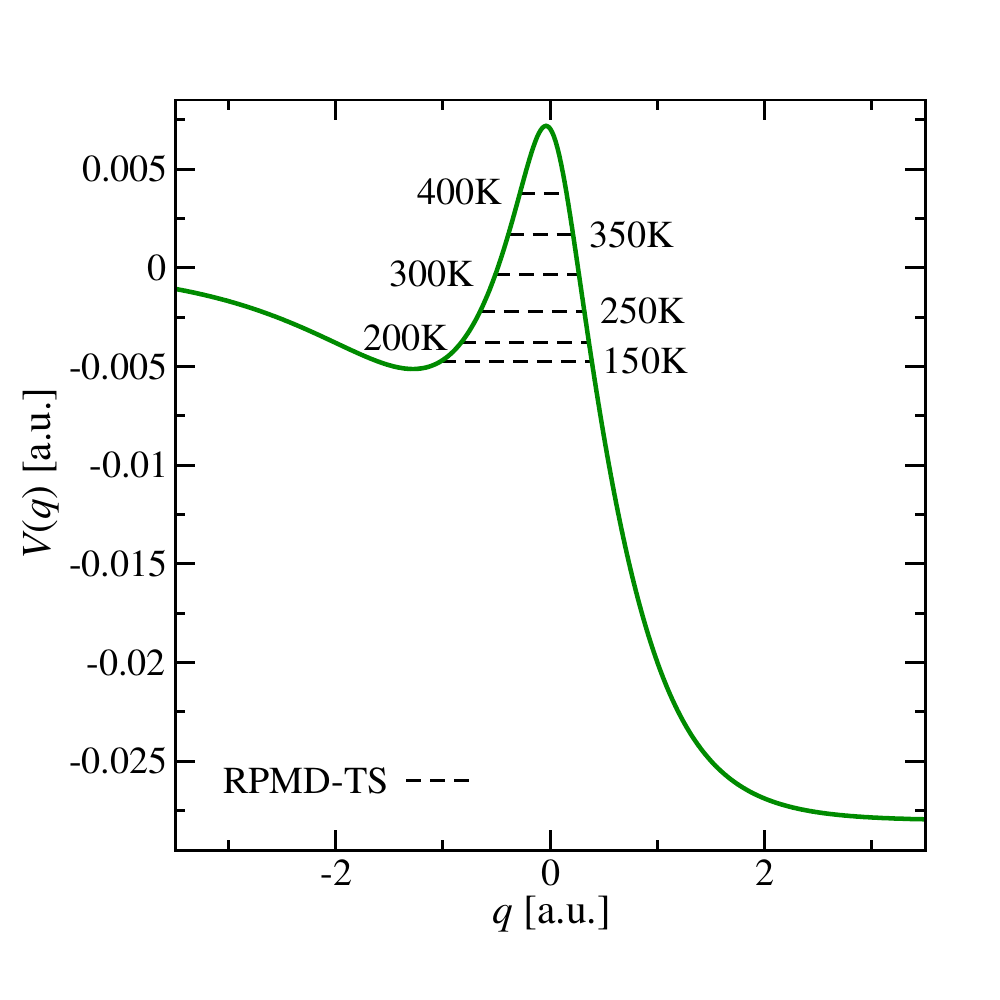}
    \vspace{-1cm}
    \caption{Plot of the one-dimensional model potential with a pre-reactive well. The ring-polymer transition states (equivalent to the semiclassical instanton trajectories) are depicted as dotted lines at a set of relevant temperatures. The failure of RPMD at low temperatures can be understood by noting that their transition states are unphysical as they correspond to energies lower than the reactant asymptote, i.e.~$E<0$. }
    \label{fig:potential}
\end{figure}

\vspace{3mm}
\noindent\textbf{Model:} 
To illustrate the breakdown of RPMD for reactions with pre-reactive minima, we consider the simplest case, a one-dimensional model, where the potential is defined as
\begin{equation}
    V(q) \!=\! \frac{V_1(q)+V_2(q)}{2}  -\sqrt{\!\left(\!\frac{V_1(q)-V_2(q)}{2}\!\right)^{\!\!2}+V^{2\phantom{/}\!\!\!}_3(q)} \,,
\end{equation}
with
\begin{subequations}
\begin{equation}
    V_1(q) = D_\mathrm{e} \left(e^{-2\alpha_1(q-q_1)}-2e^{-\alpha_1(q-q_1)}\right)
\end{equation}
\begin{equation}
    V_2(q) = D_\mathrm{e}\, e^{-2\alpha_2(q-q_2)} + \epsilon
\end{equation}
\begin{equation}
    V_3(q) = C e^{-\alpha_3^2 q^2}.
\end{equation}
\end{subequations}
The parameters used are {$D_\mathrm{e}=3.2$\,kcal\,mol$^{-1}$}, \mbox{$\alpha_1=-1$\,$a^{-1}_0$}, {$\alpha_2=1$\,$a^{-1}_0$}, {$q_1=-1.3$\,$a_0$}, {$q_2=1.24$\,$a_0$}, {$\alpha_3=1$\,$a^{-1}_0$}, $C=0.025$\,$E_{\rm h}$, and $\epsilon=-0.028$\,$E_{\rm h}$, with a mass of 2000$\,m_\mathrm{e}$. The resulting potential is shown in Fig.~\ref{fig:potential}. The barrier height, $V^\ddagger$, is approximately 7.2\,m$E_{\rm h}$ or $4.5$\,kcal\,mol$^{-1}$, and the depth of the pre-reactive well, $V(q_{\rm PRC})$, is approximately $-5.1$\,m$E_{\rm h}$ or $-3.2$\,kcal\,mol$^{-1}$. 
The barrier frequency is $\omega^\ddagger=2117$\,cm$^{-1}$ corresponding to a crossover temperature of $T_{\rm c}=485$\,K\@.
Although this potential is only one-dimensional, it contains all the key features needed to demonstrate the pathology which leads to the breakdown of RPMD  in multidimensional gas-phase reactions exhibiting pre-reactive complexes at low pressure.

\begin{figure}[t]
    \centering
    \includegraphics[width=1.0\linewidth]{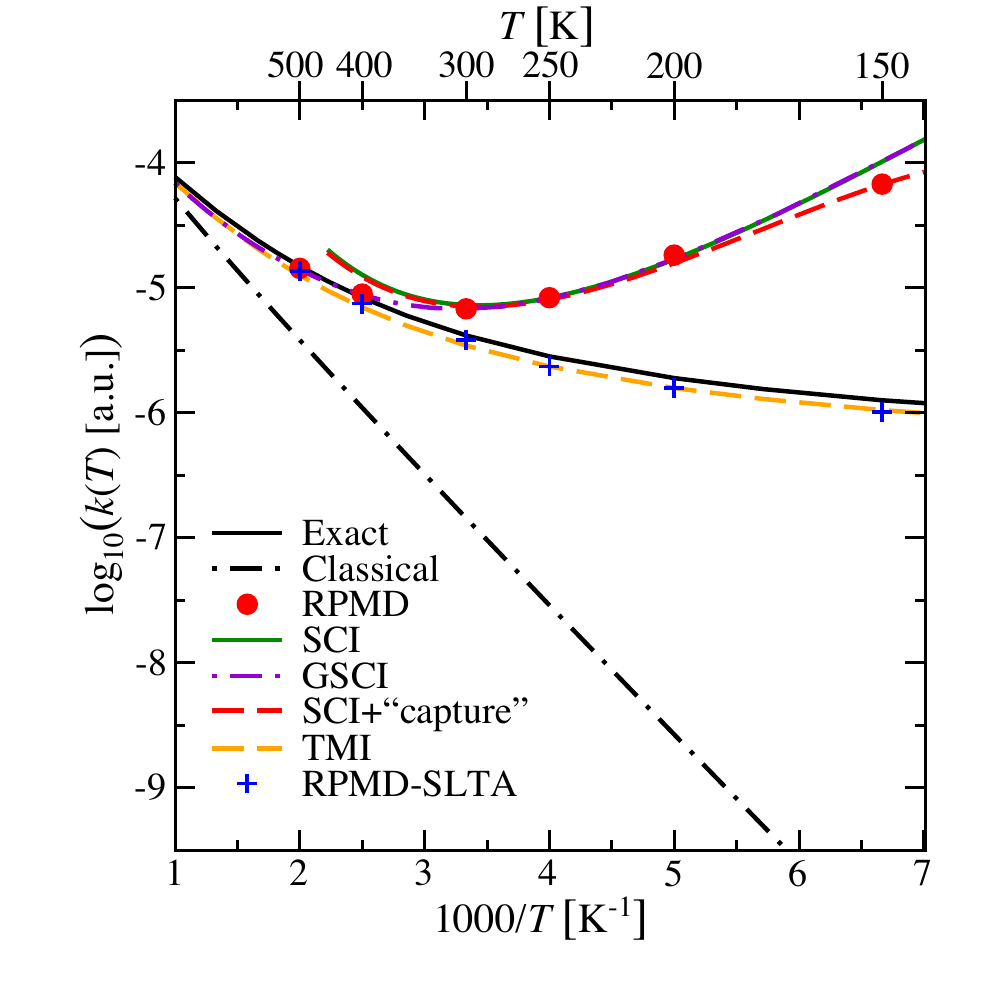}
    \caption{Arrhenius plot of the reaction rate constants in atomic units for the model potential depicted in Fig.~\ref{fig:potential}. Statistical errors in the RPMD result are smaller in all cases than the size of the circles. The RPMD-SLTA result corresponds to the RPMD result corrected using Eq.~\ref{eq:SLTA}. Note that the GSCI-SLTA rate is graphically indistinguishable from the TMI rate, and is not shown to avoid clutter.}
    \label{fig:rates}
\end{figure}

\vspace{3mm}
\noindent\textbf{Results and Discussion:}
Figure \ref{fig:rates} compares the exact quantum-mechanical rate constant (black solid line) 
for the model system as a function of inverse temperature to the rate constant predicted using RPMD (filled red circles). It is immediately clear that the RPMD rate differs significantly from the exact result. Although RPMD gives reasonable predictions at 500\,K and 400\,K, as the temperature is lowered further, the RPMD rate erroneously starts to increase while the exact rate continues to decrease. 
It is important to stress that an increase in the rate with decreasing temperature is not always unphysical, as some reactions, such as those with a submerged barrier, do show a real inverse Arrhenius effect.\cite{Song2023Submerged_Barriers}
However, in this case it is clear the behavior is pathological as it contradicts the exact result.
The error of the RPMD result is already significant at 300\,K, where it approximately a factor of 2 too large, and increases to more than an order of magnitude too large at 200\,K\@.  

In order to understand the origin of the error in RPMD, we will make use of its connection to semiclassical instanton theory (SCI).\cite{Richardson2009RPInst} As expected, the standard SCI result, shown in Fig.~\ref{fig:rates} as a green solid line, %
gives similar results to RPMD\@.
In particular, the SCI rate shows the same distinctive increase in the rate with decreasing temperature below 300\,K
and is in close agreement with RPMD at 300\,K, 250\,K, and 200\,K\@.
Near the crossover temperature, however, SCI begins to deviate from RPMD\@.
This is simply a consequence of the harmonic assumption of instanton theory, which breaks down at the crossover temperature.\cite{Affleck1981ImF}
This can, in fact, easily be overcome by using the recently derived global (uniform) semiclassical instanton theory\cite{Lawrence2024crossover} (GSCI, purple dot-dash line), which agrees closely with the RPMD result even in the crossover region. 

One can also see that  the instanton and RPMD results deviate slightly at the lowest temperatures considered (i.e.~near 150\,K). This can be explained by noting that the RPMD rate is limited by the rate at which the ring polymers arrive in the pre-reactive well, which in this one-dimensional problem is just the free-particle scattering rate $k_{\rm free}=1/\sqrt{2\pi m\beta}$. Invoking the idea of artificial thermalization that we will discuss in more detail shortly, at these low temperatures we can assume that the ring polymer will thermalize in the pre-reactive well. The RPMD rate can, therefore, be considered as a simple statistical average of the barrier transmission rate and the trivial ``capture'' rate of the well, $k_{\text{``capture''}} = k_{\rm free}$. In contrast, the instanton rate is just a local rate for crossing the barrier.    
We can, therefore, modify the instanton result to mimic the RPMD behavior using basic kinetics
 \begin{equation}
     k_{\rm SCI\text{+}\text{``capture''}} = \frac{k_{\text{``capture''}}k_{\rm SCI}}{k_{\text{``capture''}}+k_{\rm SCI}}.
\end{equation}
The resulting SCI+``capture'' rate (red dashed line) agrees closely with the RPMD result at 150\,K, leaving the other temperatures essentially unchanged.\footnote{Note that, as we will discuss later, this correction is only valid at low temperatures, where the ring polymer becomes trapped in the pre-reactive well.  We do not therefore combine it with the GSCI approach.} 
Having made the connection between the results from RPMD and instanton theory, we can analyze
the origin of the error within instanton theory, rather than considering RPMD directly.
The advantage of doing this is that instanton theory can be rigorously derived from the exact result, enabling us to give a clearer exposition of the approximations made and how they can be fixed.

The most general first-principles derivation of instanton theory uses the flux-correlation formalism.\cite{Richardson2016FirstPrinciples,Richardson2018InstReview} However, for our purpose it will be most instructive to note that, in one dimension, instanton theory can be derived starting from the exact expression for the rate constant,\cite{BellBook} 
\begin{equation}
    k(\beta)Z_r(\beta) = \frac{1}{2\pi\hbar} \int_{-\infty}^\infty P(E) \, e^{-\beta E} \, \mathrm{d}E, \label{eq:1D_rate_PE}
\end{equation}
where $P(E)$ is the transmission probability. 
The next step is to approximate the exact transmission probability with the uniform Wentzel--Kramers--Brillouin (WKB)\cite{BellBook,Kemble1935WKB,Froman1965JWKB} transmission probability
\begin{equation}
    P_{\rm WKB}(E) = \frac{1}{1+e^{W(E)/\hbar}} ,
\end{equation}
where $W(E)$ is the reduced action
\begin{equation}
    W(E) = 2 \int_{q_-(E)}^{q_+(E)} \sqrt{2m(V(q)-E)}\,dq, \label{eq:reduced_action}
\end{equation}
and $q_{\pm}$ are the turning points of the barrier defined by $V(q_{\pm})=E$. The final step is to integrate over energy asymptotically in the limit $\hbar\to0$.

\begin{figure}[t]
    \centering
    \includegraphics[width=1.0\linewidth]{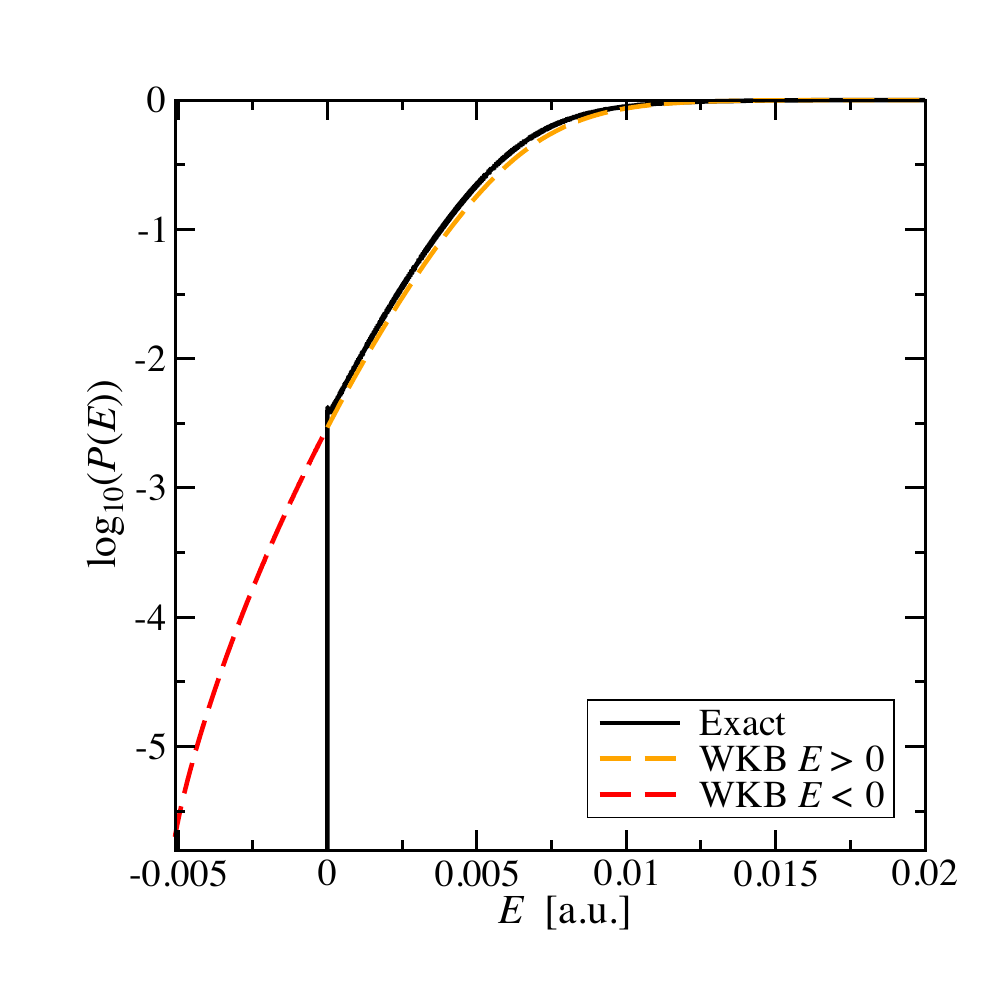}
    \caption{The exact quantum-mechanical and WKB approximations to the transmission probability.}
    \label{fig:Transmission_Probability}
\end{figure}

From this description we can immediately see the origin of the error in SCI and hence in RPMD\@. While the exact transmission probability is zero for energies below the reactant asymptote ($P(E)=0$  for $E<0$)
the semiclassical result remains non-zero for energies below the reactant asymptote but above the minimum of the pre-reactive well ($P_{\rm WKB}(E)>0$ for $V(q_{\rm PRC})<E<0$) as shown in Fig.~\ref{fig:Transmission_Probability}. 
This is because instanton theory contains only local information about the potential around the instanton path.
Hence, it does not ``know'' about the restriction on the total energy imposed by the reactant asymptote and thus uses negative energies with Boltzmann factors that unphysically increase as the temperature is lowered. 
We note that the observation that instanton theory is a local theory is not new, and the need to account for this fact in systems exhibiting pre-reactive complexes has been made before in e.g.~Refs.~\citenum{Faraday2016fundamentals,Alvarez2016NH3+H2,DoSTMI,JoeFaraday}.
However, here we will use these observations to explain the failure of RPMD and provide a simple correction.
RPMD is fundamentally a thermal theory. Hence, to understand the error in more detail, we need to analyze how the error in the energy domain manifests in the resulting thermal expression. For this purpose we consider the standard SCI expression (which was plotted in Fig.~\ref{fig:rates})
\begin{equation}
    k_{\rm SCI}(\beta)Z_r(\beta) = \frac{1}{2\pi\hbar} \sqrt{\frac{2\pi\hbar}{W''(E^\star)}} \, e^{-W(E^\star)/\hbar-\beta E^\star} ,
\end{equation}
where $E^\star(\beta)$ is defined by $W'(E^\star)=-\beta\hbar$.  This expression is obtained by approximating the integrand in Eq.~\ref{eq:1D_rate_PE} by a Gaussian. Clearly, $E^\star<0$ corresponds to an ``unphysical'' instanton path.
Figure~\ref{fig:potential} visualizes the instanton paths at various temperatures showing their corresponding energies. Comparing with Fig.~\ref{fig:rates} thus explains the significant errors at 150\,K, 200\,K and 250\,K as arising from spurious contribution to the rate from energies which are not physically accessible.  
One might think that the problem would go away as soon as $E^\star\geq0$. However, 
even when $E^\star=0$, the absence of the lower bound in the integral will result in  overestimating the true rate by approximately a factor of two. This is because we are effectively approximating the integral over energy by the area of the entire Gaussian, rather than just half of it. %

The clearest demonstration that you understand a problem is that you can fix it. With this in mind we note that this analysis clearly implies that in one dimension we can fix the error of instanton theory by simply integrating the WKB transmission probability over energy numerically with an appropriate lower limit
\begin{equation}
    k_{\rm TMI}(\beta)Z_r(\beta) = \frac{1}{2\pi\hbar} \int_{0}^\infty P_{\rm WKB}(E) \, e^{-\beta E} \,\mathrm{d}E. \label{eq:TMI}
\end{equation}
Correcting thermal instanton theory in this manner, in cases where the local distribution is not thermal, has been discussed previously where it was referred to as the thermalized microcanonical instanton theory (TMI).\cite{Faraday,DoSTMI,JoeFaraday}
The resulting rate constant is plotted in Fig.~\ref{fig:rates} (orange dashed line) and shows good agreement with the exact result at all temperatures, giving a clear graphical confirmation of our analysis.\footnote{We note here that using more accurate semiclassical approximations to $P(E)$ involving higher derivatives of the potential, such as those discussed in Refs.~\citenum{Upadhyayula2024hbar2corrections,Pollak2024hbar4corrections,Pollak2024Perspective}, would of course give even better agreement with the exact result.} 

Before discussing how to overcome the problem in multidimensional systems, we return to discuss how our analysis can be translated to the language of RPMD\@. 
At the simplest level, the connection between the optimum RPMD transition-state theory and instanton theory\cite{Richardson2009RPInst} implies that the breakdown of RPMD is caused by the transition-state ensemble including paths that correspond to unphysical energies.
Ultimately, we will argue that this can be explained as arising from a general phenomenon of RPMD that we term ``artificial thermalization.''
To understand the origin of artificial thermalization, 
it is helpful to separate the dynamics of the ring polymer into the motion of its centroid and internal degrees of freedom.
The large ($n\to\infty$) number of internal degrees of freedom can then be viewed as an effective ``bath'' coupled to the centroid coordinate. 
This perspective allows us to identify two types of artificial thermalization that occur in RPMD, one associated with the effective renormalization of the potential and the other associated with the friction experienced by the centroid.

\begin{figure}[t]
    \centering
    \includegraphics[width=1.0\linewidth]{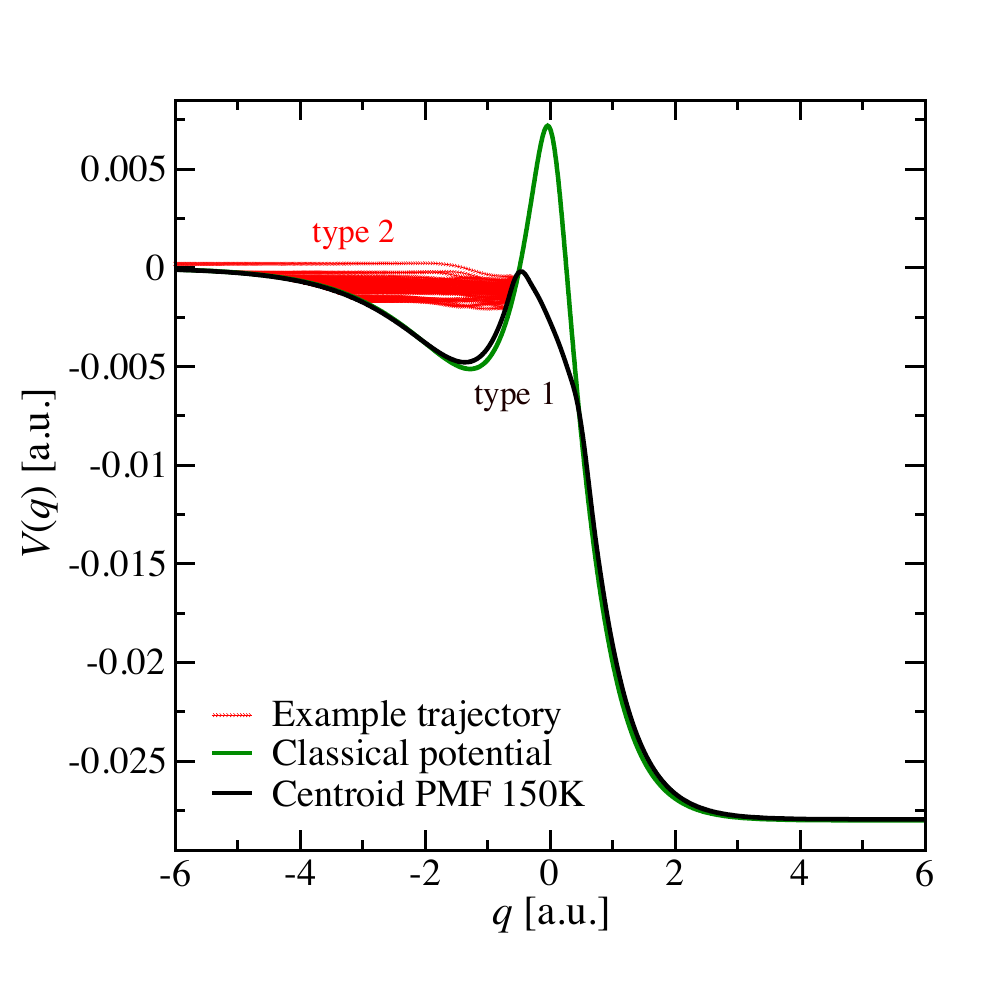}
    \caption{Illustration of the two types of artificial thermalization viewed in terms of the centroid dynamics. The red line shows the centroid energy (kinetic energy of centroid plus potential of mean force) of a typical trajectory that thermalizes (becomes trapped) in the pre-reactive well at 150\,K.
    The extremely low height of the PMF barrier is caused by contamination of quantum states with energies lower than the reactant asymptote, leading to an unphysically high prediction of the rate at low temperature.
    }
    \label{fig:relax}
\end{figure}

First, for a fixed location of the centroid, the internal modes of the ring polymer sample the potential in a manner that is consistent with the \emph{local} thermal density.
Crucially, this means the potential of mean force (PMF) in the barrier region contains contributions from quantum states with energies both above and below the reactant asymptote.
Although of course the PMF is supposed to be lower than the classical barrier so as to mimic the effect of tunneling, in this case it is lowered too much, as depicted in Fig.~\ref{fig:relax}.
As the rate depends exponentially on the barrier height, this type of artificial thermalization is the dominant effect that leads to the breakdown of the RPMD rate. %

The second type of artificial thermalization is the thermalization of the centroid.
In contrast to centroid molecular dynamics (CMD),\cite{Voth+Chandler+Miller,Jang1999CMDa,Jang1999CMDb,Geva2001CMD} in RPMD the centroid is not adiabatically separated from the internal modes, and feels  additional effective frictional and fluctuating forces on top of the PMF\@. 
These forces are crucial for the improved accuracy that RPMD exhibits over CMD in thermal tunneling through asymmetric barriers.\cite{Richardson2009RPInst} 
However, from the perspective of the centroid these additional forces cause thermalization, as depicted by the red line in Fig.~\ref{fig:relax} that shows a typical trajectory at 150\,K relaxing into the pre-reactive well and hence becoming ``trapped'' for a long time.
The time scale for this thermalization is shorter at low temperatures, because the larger radius of the ring polymer samples more anharmonicity and, hence, leads to a stronger coupling between the centroid and the internal modes.

One might still ask, %
given that RPMD is a fully dynamical theory and can explore the global potential, why does the existence of the reactant asymptote fail to leave its mark, as it does in the quantum-mechanical theory?
In particular, one might assume that in calculating the transmission coefficient, recrossing trajectories may compensate for the low barrier of the PMF\@.
However, to the extent that the reactant asymptote constrains the ring-polymer energy it only constrains the energy of the centroid, because in the $n\to\infty$ limit the total energy of the ring polymer is unbounded.
Crucially, because the PMF is a thermal average, the classical energy of the centroid does not correspond directly to the quantum energy.  So any constraint on the centroid energy cannot not mimic the true effect of the reactant asymptote on the tunneling rate. Furthermore, at the low temperatures we are considering, the second type of artificial thermalization results in the energy of the centroid in RPMD rapidly thermalizing. So practically there is no constraint on the energy at all. This explains why the full RPMD rate can be understood quantitatively in terms of the simple instanton/optimum RPMD transition-state theory analysis given earlier. 

At this point it is helpful to make connection to the previous studies that have noticed issues when applying RPMD to systems with pre-reactive complexes.\cite{delMazoSevillano2023PRC_RPMD_(H2CO)2+OH,delMazo2021OH+H2CO,Hashimoto2023RPMDNH3H2} These studies observed two key phenomena: first, that the rate exhibited a spurious increase at low temperatures; second, that the RPMD trajectories often became trapped for very long times in the pre-reactive complex. On the basis of the preceding analysis we can now attribute these two issues to the two different types of artificial thermalization present in RPMD\@. The trapping occurs because of the second type of thermalization: the thermalization of the centroid into the pre-reactive well, which makes re-dissociation statistically very unlikely. 
However, the preceding analysis shows that this is not the principle effect that leads to a breakdown of RPMD rate theory.
Rather, this is caused by the first type of artificial thermalization: the internal modes of the ring polymer sampling all locally thermally accessible states. This is inherent to all imaginary-time path-integral methods and leads to contamination of the rate by negative-energy states with unphysically large Boltzmann factors.

Hence, while CMD does not suffer from the second type of artificial thermalization, and so CMD trajectories cannot become trapped in the pre-reactive well,  it would still exhibit spuriously large rates at low temperatures due to the first type of artificial thermalization.
In fact,
the CMD rate is actually guaranteed to be even larger than the RPMD rate for this system.  
This is because in one dimension the CMD rate,\cite{Voth+Chandler+Miller,Jang1999CMDa,Jang1999CMDb,Geva2001CMD} which is just the classical rate calculated on the centroid PMF, is trivially always an upper bound to the RPMD rate. %

In earlier studies,\cite{delMazo2021OH+H2CO,delMazoSevillano2023PRC_RPMD_(H2CO)2+OH} it was suggested the trapping was caused by the ``spurious resonances'' exhibited by RPMD vibrational spectra.\cite{Witt2009spectroscopy,Rossi2014resonance} These spurious resonances occur when the frequencies of the internal modes of the ring polymer come close to the physical frequencies of the system, resulting in resonant energy transfer between the physical and internal ring-polymer degrees of freedom.
There are several alternative imaginary-time path-integral methods that do not exhibit spurious resonances in their vibrational spectra, notably CMD,\cite{Jang1999CMDa,Jang1999CMDb,Musil2022PIGS} quasi-centroid MD (QCMD),\cite{Trenins2019QCMD,Trenins2022QCMD,Benson2020water,Fletcher2021fQCMD,Lawrence2023fQCMD,Lieberherr2023fQCMDPolariton,Limbu2025hfQCMD} and thermostatted-RPMD (TRPMD).\cite{Rossi2014resonance,Litman2020FaradayTRPMD} 
Crucially, however, it is not the ``resonance'' that is important here, but rather that the internal modes and the centroid are coupled at all. Damping of the resonance does not remove this coupling\footnote{In fact they simply smear the resonance peaks in the vibrational spectrum.\cite{Benson2020water}} and so TRPMD trajectories still become trapped by the second type of artificial thermalization (which explains the observations of Ref.~\citenum{delMazoSevillano2023PRC_RPMD_(H2CO)2+OH}).  
Hence, while the spurious resonances and trapping are related, they are not  the same phenomenon.
In contrast, CMD and QCMD, which employ a true adiabatic separation between the centroid/quasi-centroid and the internal modes, will not exhibit the second type of artificial thermalization. However, all these methods still exhibit the first type of artificial thermalization and hence will predict unphysical rates for systems with pre-reactive complexes.

At this point, it is important to stress that there is no simple fix to the ring-polymer dynamics that will avoid the artificial thermalization. This is because the connection between the quantum-mechanical energy and the approximate dynamics of the ring polymer is indirect. 
In one dimension, Eq.~\ref{eq:TMI} indicates that modifying the potential to simply ``fill in'' the well would likely significantly improve the RPMD result. However, this approach cannot be generalized to multiple dimensions as it is the total energy and not the energy along the reaction coordinate that should be constrained. 
A natural approach to fix the problem would be to start with Matsubara dynamics\cite{Hele2015Matsubara,Althorpe2021Matsubara,Althorpe2024ReviewVibrationalSpectra} and derive a new alternative to RPMD and CMD\@. Unfortunately, one would almost certainly need to incorporate terms with complex phase factors, leading to the infamous sign problem.  
However, this is not to say that there are no practical ways to overcome the problem.

The above discussion of the breakdown of RPMD is just as valid in multiple dimensions as it is in one.  However, fixing the issue is more challenging when there is more than one degree of freedom. 
The generalization of the transmission probability to multiple dimensions is the cumulative reaction probability, $N(E)$, which can be thought of as the number of reactive channels at a given energy.\cite{Miller1993QTST} Accurately estimating $N(E)$ is fundamentally more difficult than calculating $k(\beta)$, reflecting the greater amount of information in a microcanonical theory than is available in the canonical (thermal) ensemble,\footnote{This is equivalent to saying that the inverse Laplace transform is ill-posed.} and this difficulty only increases in multiple dimensions. 

Due to the global nature of energy eigenstates, the exact $N(E)$ is a global quantity.
However, as with $P_{\rm WKB}(E)$ in one dimension, practical approximations to $N(E)$ are always local. %
These approximate $N_{\rm local}(E)$ can be formally defined as the inverse Laplace transform of a local Boltzmann-weighted flux, $\Phi_{\rm local}(\beta)$, according to
\begin{equation} \label{eq:Phi_local}
    \Phi_{\rm local}(\beta) =  c_{\rm fs}(t_{\rm p};\beta) = \frac{1}{2\pi\hbar}\int_{-\infty}^{\infty} N_{\rm local}(E) \, e^{-\beta E} \, \mathrm{d}E ,
\end{equation}
where $c_{\rm fs}(t_{\rm p};\beta)$ is a flux--side correlation function suitable for the local rate process of interest and $t_{\rm p}$ is the plateau time.\footnote{Formally, this relies on $\Phi_{\rm local}(\beta)$ being well defined for all $\beta>0$. However, for practical purposes, at a given energy, $N_{\rm local}(E)$ is only sensitive to a restricted range of $\beta$.} The local thermal flux is related to a local thermal rate constant via the corresponding reactant partition function by $\Phi_{\rm local}(\beta)=k_{\rm local}(\beta)Z_r(\beta)$.

Multiple approaches to calculating microcanonical rates exist,\cite{Marcus1952RRKM,Miller1987unimolecular,Truhlar1996TST,Muga2004CAPs,Shan2019SCTSTReview} and we will not give a detailed review of each of these approaches here. One of the simplest uses the stationary-phase approximation to the inverse Laplace transform\cite{Forst1971InverseLaplaceTransform,DoSTMI,Tao2020microcanonical} to give
\begin{equation}
\begin{aligned}
    N^{\rm SPA}_{\rm local}(E) &=  \hbar \,\sqrt{-\frac{2\pi}{\mathcal{E}'(\beta_{\rm sp})}} \, e^{\,E\beta_{\rm sp}(E)} \, \Phi_{\rm local}\big(\beta_{\rm sp}(E)\big) ,
    \end{aligned}
\end{equation}
where $\beta_{\rm sp}$ is defined %
as the solution of $E=\mathcal{E}(\beta)$ and
\begin{equation}
    \mathcal{E}(\beta) = - \frac{\mathrm{d} \ln \Phi_{\rm local}(\beta)}{\mathrm{d}\beta} 
\end{equation}
is an effective average reaction energy.
 This approach can be combined with any approximate theory for $\Phi_{\rm local}(\beta)$. The only requirement is that $\mathcal{E}(\beta)$ is a monotonically decreasing function of $\beta$, which simply means that the thermal flux must be consistent with a positive $N_{\rm local}(E)$. Practically, however, calculating $N^{\rm SPA}_{\rm local}(E)$ is far more challenging for numerical methods such as RPMD that exhibit statistical error, than for analytical methods like instanton theory.\footnote{This is not only because of the statistical error, but also because such theories typically do not give ready access to absolute quantities such as $Z(\beta)$ or $\Phi(\beta)$.}

On this basis we now consider an approach that avoids needing to directly calculate any $N_{\rm local}(E)$.
When modeling bimolecular reactions with pre-reactive complexes, one must in general compute local rates associated with both the capture process (formation of the pre-reactive complex) and the barrier crossing, and then account for which is the limiting process.  However, if one assumes the barrier crossing is always the rate-limiting step then the zero-pressure rate constant can simply be expressed as
\begin{subequations}
\begin{equation}
   k(\beta) = \frac{\Phi_{\rm barrier}(\beta;E_{\rm min})}{Z_r(\beta)} 
\end{equation}
    \begin{equation}
    \Phi_{\rm barrier}(\beta;E_{\rm min})= \frac{1}{2\pi\hbar} \int_{E_{\rm min}}^\infty N_{\rm barrier}(E) \, e^{-\beta E} \,\mathrm{d}E ,\label{eq:Integral_with_Emin_limit}
\end{equation}
\end{subequations}
where $E_{\rm min}$ is the minimum energy that the reactants can have\footnote{This is equivalent to assuming that $N_{\rm capture}(E)\gg N_{\rm barrier}(E)$ for $E\geq E_{\rm min}$ and $N_{\rm capture}(E)=0$ for $E<E_{\rm min}$, i.e.,~$N_{\rm capture}(E)$ is modeled as turning on sharply at the minimum energy.} 
and $N_\mathrm{barrier}(E)$ is the local approximation to $N(E)$ based on information in the barrier region only.
In this case, it is possible to directly approximate the effect on $\Phi_{\rm barrier}(\beta)$ of the modified lower bound in the Laplace transform, $E_{\rm min}$, and hence avoid calculating $N_{\rm barrier}(E)$. 
As we show in the supplementary material, with a careful asymptotic analysis (making use of Bleistein's method for deriving uniform asymptotic series\cite{Bleistein1966Uniform,RWong1989UniformAsymptotics}) we can derive the following ``shifted Laplace transform approximation'' (SLTA)
\begin{subequations}  \label{eq:SLTA}
    \begin{equation}
     k^{\rm SLTA}(\beta;E_{\rm min})= k_{\rm barrier}(\beta)\,\theta^{\rm SLTA}(\beta;E_{\rm min})
\end{equation}
\begin{equation}
\begin{aligned}
    \theta^{\rm SLTA}(\beta;E_{\rm min})=\Bigg[&\frac{1}{2}  \mathrm{erfc}\left(\mathrm{sgn}(\beta_{\rm min}-\beta)\sqrt{\Delta R}\right)\\ &+ \frac{e^{-\Delta R}}{\sqrt{4\pi\Delta R}} \mathrm{sgn}(\beta_{\rm min}-\beta)\\ &+\frac{e^{-\Delta R}}{\sqrt{-2\pi\mathcal{E}'(\beta_{\rm min})}(\beta-\beta_{\rm min})}\Bigg] ,
\end{aligned}
\end{equation}
\end{subequations}
where $\beta_{\rm min}=\beta_{\rm sp}(E_{\rm min})$ can be found by solving
\begin{equation}
    \mathcal{E}(\beta_{\rm min}) = E_{\rm min}
\end{equation}
and
\begin{equation}
    \Delta R = (\beta - \beta_{\rm min})E_{\rm min} - \ln \Phi(\beta_{\rm min}) + \ln \Phi(\beta).
\end{equation}
This has two key advantages over calculating $N^{\rm SPA}_{\rm barrier}(E)$ directly and integrating Eq.~\ref{eq:Integral_with_Emin_limit}: first, it is guaranteed to recover the original $k_\mathrm{barrier}(\beta)$ in the limit that ${\beta_{\rm min}\to\infty}$, e.g.~as the depth of the pre-reactive well goes to zero; second, it is less computationally demanding, as it requires calculations at fewer values of $\beta$ and one only needs $\mathcal{E}'(\beta_{\rm min})$ [which is related to the second derivative of $\ln \Phi(\beta)$ and can be expensive to converge for methods like RPMD] at a single value of $\beta$.   

As an illustrative example, in Fig.~\ref{fig:rates} we also include the RPMD rate corrected using Eq.~\ref{eq:SLTA} (RPMD-SLTA, blue crosses). The correction agrees almost perfectly with the TMI result and thus approximates the exact rate well. This is unsurprising as (although to avoid clutter we do not plot it) the GSCI-SLTA result is also graphically indistinguishable from the TMI result. 
For RPMD-SLTA, we find that $\beta_{\rm min}=1088\,E_{\rm h}^{-1}$ (i.e.~a temperature of $292\,{\rm K}$), which corresponds roughly to the temperature at which the instanton energy $E_\mathrm{I}(\beta)=E_{\rm min}=0$. Note, in multidimensional cases this generalizes to $E_\mathrm{I}(\beta)+E_{\rm vib}(\beta)+E_{\rm rot}(\beta)\approx E_{\rm min}$, where $E_{\rm vib}(\beta)$ and $E_{\rm rot}(\beta)$ are effective vibrational and rotational energies arising from the orthogonal degrees of freedom.

For this simple one-dimensional system, obtaining the RPMD-SLTA rate is a trivial post-processing step that simply requires multiplying the local barrier-crossing rate $k_{\rm barrier}(\beta)$ by the reactant partition function per unit length, $Z_r=\sqrt{m/(2\pi\beta\hbar^2)}$, and then fitting a cubic spline through a small number of resulting $\Phi_{\rm barrier}(\beta)$ values. 
Crucially, the calculation of the barrier crossing rate is much less computationally demanding than the full RPMD rate, as one can use a local committor about the barrier that treats trajectories that become ``trapped'' in the well as having returned to the reactants.

To calculate the RPMD-SLTA rate in multiple dimensions would require a couple of extra steps, as one does not typically know the reactant partition function analytically. 
Fortunately, one does not need to compute the absolute $Z_r$; instead one can in principle obtain all necessary data by simply computing the internal energy of the reactants, $U_r(\beta)=-\mathrm{d}\ln Z_r/\mathrm{d}\beta$, e.g.~using a virial estimator,\cite{TuckermanBook} at each value of $\beta$. The calculation of the RPMD-SLTA rate can then be carried out following this simple post-processing recipe: (1) Cubic spline the $\ln k_{\rm local}(\beta)$ data; (2) Differentiate the spline %
to give $\mathcal{E}(\beta)=U_r(\beta)-\mathrm{d}\ln k_{\rm local}/\mathrm{d}\beta$; (3) Spline the resulting $\mathcal{E}(\beta)$ data and solve to find $\mathcal{E}(\beta_{\rm min})=E_{\rm min}$; (4) Differentiate the spline to obtain $\mathcal{E}'(\beta_{\rm min})$;   (5) Integrate the spline  to give $\ln \Phi_{\rm local}(\beta) - \ln \Phi_{\rm local}(\beta_{\rm min})$ and hence $\Delta R(\beta)$; (6) Combine $\beta_{\rm min}$ and $\mathcal{E}'(\beta_{\rm min})$ with the $\Delta R(\beta)$ and $k_{\rm local}(\beta)$ data to obtain the RPMD-SLTA rate according to Eq.~\ref{eq:SLTA}.

Finally we note that, in general, modeling real chemical reactions is not as simple as just integrating over $N_{\rm barrier}(E)$ with an appropriate lower bound.
For a reaction involving a stable pre-reactive complex, one must consider whether the number of channels for the formation of the pre-reactive complex or the barrier crossing is larger.
In the zero-pressure limit, assuming these two processes are statistically independent, a more accurate estimation of the total cumulative reaction probability is then given by\cite{Light1967FaradayStatisticalRate,Miller1976rate}
\begin{equation}
    N_{\rm tot}(E) = \frac{N_{\rm barrier}(E)N_{\rm capture}(E)}{N_{\rm barrier}(E)+N_{\rm capture}(E)} , \label{eq:low_pressure_NE}
\end{equation}
where $N_{\rm capture}(E)$ is the number of channels for the formation of the pre-reactive complex. 
Of course not all reactions are in either the low- or high-pressure limits.
In general the effect of collisional energy transfer can be incorporated to calculate pressure-dependent rate constants by combining the $N_{\rm local}(E)$ with rates for collisional energy transfer in a general master equation.\cite{Glowacki2012MESMER,Georgievskii2013MESS,Zhang2022TUMME,MultiWell}

\vspace{3mm}
\noindent\textbf{Conclusions:}
We have demonstrated that, despite its many successes, great care must be taken when applying RPMD rate theory to systems where the assumption of local thermal equilibrium in the barrier region is not valid. 
By considering a simple model of a gas-phase reaction with a pre-reactive minimum, we have shown how RPMD rate theory can break down at low temperatures in the low-pressure limit.\footnote{Given that RPMD obeys detailed balance, it is trivial to see it will also break down for endothermic reactions with post-reactive minima.}

Using the connection of RPMD rate theory and instanton theory, we have shown that this breakdown can be understood in terms of contamination of the tunneling rate by quantum states with energies below the reactant asymptote.
Indications of issues with RPMD for such reactions were already present in the literature, with observations both of spuriously large rates and the trapping of trajectories in the pre-reactive well.\cite{delMazoSevillano2023PRC_RPMD_(H2CO)2+OH,delMazo2021OH+H2CO,Hashimoto2023RPMDNH3H2}
Here, we have explained these issues in terms of two types of artificial thermalization that affect RPMD.
First, the 
ring polymer always samples all quantum states weighted by their local thermal density.
Hence, in a system with a pre-reactive minimum, tunneling through the barrier is contaminated by states with unphysically low energies and large Boltzmann factors. %
The second type of artificial thermalization, which explains the observed trapping, is the classical thermalization of the centroid motion caused by coupling to the internal modes of the ring polymer.

Unfortunately, this breakdown is very difficult to overcome by simply modifying the ring-polymer dynamics.
Previous interpretations of the issue in terms of the ``spurious resonances'' of RPMD\cite{delMazo2021OH+H2CO,delMazoSevillano2023PRC_RPMD_(H2CO)2+OH} would imply that methods such as  CMD,\cite{Jang1999CMDa,Jang1999CMDb,Musil2022PIGS} QCMD,\cite{Trenins2019QCMD,Trenins2022QCMD,Benson2020water,Fletcher2021fQCMD,Lawrence2023fQCMD,Lieberherr2023fQCMDPolariton,Limbu2025hfQCMD} or TRPMD\cite{Rossi2014resonance,Litman2020FaradayTRPMD} would fix the problem. However, the present analysis clearly demonstrates that each of these methods would still suffer at least from the first type of artificial thermalization and hence exhibit spuriously large tunneling rates at low temperatures. 

More generally, our analysis shows that RPMD will fail to describe any process where it is simultaneously the case that quantum effects are important and local thermalization (e.g.~vibrational relaxation) rather than spatial diffusion is the slowest process.
This general principle can also be used to understand the failure of RPMD to describe other phenomena. For example, RPMD does not capture resonances in the rate for systems exhibiting polaritonic vibrational strong coupling, which is attributed to dynamics dominated by energy diffusion.\cite{Fiechter2023HowQuantumPolaritonRPMD,Lindoy2023polariton}
This principle extends beyond simply vibrational relaxation and can also be used to rationalize the difficulty of generalizing RPMD to treat electronically nonadiabatic systems particularly in the famous Marcus inverted regime.\cite{TimMastersarxiv,Menzeleev2011ET,Shushkov2013instanton,Kretchmer2016KCRPMD,Kretchmer2018KCRPMD,Lawrence2018Wolynes,Tao2018isomorphic,Lawrence2019isoRPMD,Lawrence2019ET,Lawrence2020Improved,NRPMDChapter}
 Ultimately, this should serve as an important reminder that RPMD is only accurate at simulating long-time dynamics provided the process in question can be accurately described  via the short-time behavior of a linear-response correlation function.
 This highlights the need for an abundance of caution when using approaches that attempt to extract microcanonical information directly from RPMD trajectories in systems where quantization, beyond zero-point energy, is important.\cite{Tao2020microcanonical,Marjollet2020RPMD,Marjollet2022NERPMD_H+CH4_F+CHD3}

Although the breakdown of RPMD cannot be avoided by modifying the dynamics, we have illustrated how accurate results can still be obtained by extracting microcanonical information from local thermal calculations.
We have emphasized that the calculation of local thermal RPMD rates has the practical advantage that they do not suffer from numerical issues associated with trapping of trajectories.\cite{delMazoSevillano2023PRC_RPMD_(H2CO)2+OH,delMazo2021OH+H2CO,Hashimoto2023RPMDNH3H2} 
Further, under the assumption that the barrier crossing remains the rate-limiting process at all energies, we have introduced a simple procedure for directly estimating the thermal rate in systems with pre-reactive complexes.
This direct approach modifies the local thermal barrier-crossing rate to account for the constraint on the minimum energy imposed by the reactant asymptote. 
This avoids needing to calculate $N_{\rm barrier}(E)$ from the stationary-phase approximation to the inverse Laplace transform,\cite{Forst1971InverseLaplaceTransform,DoSTMI,Tao2020microcanonical} which can be costly for methods such as RPMD that suffer from statistical error.

However, in general, it may still be necessary to calculate $N(E)$. In these cases, to avoid the statistical error associated with RPMD while retaining its ability to capture both anharmonicity and deep tunneling, it would be desirable to have a version of instanton theory that can capture the influence of anharmonic modes on reaction rates.
Our perturbatively corrected ring-polymer instanton theory (RPI+PC), which was originally developed to calculate quantitatively accurate tunneling splittings in molecular systems,\cite{AnharmInst}
and which has been recently extended to calculating reaction rates,\cite{PCIRTarxiv}
may offer the ultimate solution for simulating reactions with pre-reactive complexes.

\vspace{3mm}
\noindent\textbf{Methods:} RPMD calculations were performed using in-house code. All simulations used a time step of 20 atomic time units and $n=256$ ring-polymer beads. The calculations at 200--500\,K were performed using the Bennet--Chandler approach\cite{Bennett1977TST,Chandler1978TST} with a centroid dividing surface, whose optimal location was roughly determined using a short thermodynamic integration. Converged values of the PMF at this location were then obtained using thermodynamic integration from $q_{\rm c}=-12\,{a}_0$ at Gaussian quadrature points with a total of 60 steps.
As the centroid PMF at 150\,K was found to have a submerged barrier, the trajectories were initialized in the reactant asymptote with the initial centroid position $q_{\rm c}=-12\,{a}_0$.
In some cases, very long runs were required to properly converge the transmission coefficient;
trajectories were only terminated
once the centroid reached $q_{\rm c}<-14\,{a}_0$ or $q_{\rm c}>2\,{a}_0$. %
At each temperature, 10,000 trajectories were used to calculate the transmission coefficient. 
Note that as RPMD is independent of the choice of dividing surface, identical results would in principle be obtained by simply propagating ring polymers from the reactant asymptote, although in practice, this would be more difficult to converge.

Exact quantum rates were calculated using the log-derivative scattering method\cite{Johnson1973logderivative,Mano1986logderivative} and integrating Eq.~\ref{eq:1D_rate_PE} numerically. Instanton rates were obtained by calculating $W(E)$ on a grid of energies using numerical integration of Eq.~\ref{eq:reduced_action} and its derivatives with respect to energy.

\begin{acknowledgement}
JEL was supported by an ETH Zurich Postdoctoral Fellowship and an Independent Postdoctoral Fellowship at the Simons Center for Computational Physical Chemistry, under a grant from the Simons Foundation (839534, MT).
\end{acknowledgement}

{\footnotesize
\renewcommand{\refname}{\normalsize References}
\setlength{\bibsep}{0pt}      %
\setlength{\itemsep}{0pt}     %
\setlength{\parskip}{0pt}     %
\setlength{\parsep}{0pt}      %
\bibliography{references,extra_refs} %
}

\end{document}


\maketitle

\renewcommand{\thepage}{S\arabic{page}}
\renewcommand{\theequation}{S\arabic{equation}}
\renewcommand{\thefigure}{S\arabic{figure}}
\renewcommand{\thetable}{S\arabic{table}}
\renewcommand{\thesection}{S\arabic{section}}
\renewcommand{\thesubsection}{S\arabic{section}.\arabic{subsection}}

\section{Stationary-Phase Approximation to Inverse Laplace Transform and Shifting the Lower Limit of Integration}
We begin by recapping the basic theory for the calculation of the inverse Laplace transform by the stationary-phase approximation.\cite{Forst1971InverseLaplaceTransform} Our starting point is the definition
\begin{equation}
  k(\beta)\,Z_r(\beta) = \Phi(\beta) = \frac{1}{2\pi\hbar}\int_{-\infty}^\infty N(E) \, e^{-\beta E} \, \mathrm{d}E ,
\end{equation}
where $N(E)$ is the ``cumulative reaction probability'' or the microcanonical number of reactive states.
Hence, formally we have that the cumulative reaction probability can be recovered using the Bromwich integral
\begin{equation}
    N(E) = \hbar \int_{-\infty}^{\infty} e^{-iv E} \, \Phi(-iv) \,\mathrm{d}v,
\end{equation}
where $v=i\beta$.
Introducing a perturbative parameter $\lambda$, we can integrate this by stationary phase:
\begin{equation}
\begin{aligned}
    N(E) &= \hbar \int_{-\infty}^{\infty} e^{-(iv E-\ln \Phi(-iv))/\lambda} \,\mathrm{d}v \\
    &\sim \hbar \sqrt{2\pi\lambda}\,\Bigg(\!-\frac{\mathrm{d}\mathcal{E}}{\mathrm{d}\beta}\Bigg)^{-\nicefrac{1}{2}}_{\beta=\beta_{\rm sp}} \, e^{\,(\beta_{\rm sp}E + \ln \Phi(\beta_{\rm sp}))/\lambda} \,\text{ as }\,\lambda\to0 ,
    \end{aligned}
\end{equation}
where 
\begin{equation}
    \mathcal{E}(\beta) = - \frac{\mathrm{d} \ln \Phi(\beta)}{\mathrm{d}\beta} \label{eq:Energy_Definition}
\end{equation}
and $\beta_{\rm sp}$ is defined implicitly by
\begin{equation}
  E = \mathcal{E}(\beta_{\rm sp}),
\end{equation}
i.e.~$\beta_{\rm sp}(E)$ is the inverse function $\beta_{\rm sp}(E)=\mathcal{E}^{-1}(E)$. Setting $\lambda=1$ gives the stationary-phase approximation to the inverse Laplace transform
\begin{equation}
\begin{aligned}
    N^{\rm SPA}(E) &=  \hbar \sqrt{2\pi}\,\Bigg(\!-\!\frac{\mathrm{d}\mathcal{E}}{\mathrm{d}\beta}\Bigg)^{-\nicefrac{1}{2}}_{\beta=\beta_{\rm sp}} \, e^{\,E\beta_{\rm sp}(E)}  \,\Phi\big(\beta_{\rm sp}(E)\big) .
    \end{aligned}
\end{equation}

The accuracy of the approximation can be determined by numerical integration to recover an approximation to $\Phi(\beta)$,
\begin{equation} \label{eq:LT}
    \Phi(\beta) \approx \frac{1}{2\pi\lambda} \int_{-\infty}^\infty  \sqrt{2\pi\lambda}\,\Bigg(\!-\!\frac{\mathrm{d}\mathcal{E}}{\mathrm{d}\beta}\Bigg)^{-\nicefrac{1}{2}}_{\beta=\beta_{\rm sp}} \, e^{-((\beta-\beta_{\rm sp})E - \ln \Phi(\beta_{\rm sp}))/\lambda} \, \mathrm{d} E,
\end{equation}
after setting $\lambda=1$.
However, regardless of the accuracy of this approximation, performing the Laplace transform integral asymptotically by steepest descent is self-consistent, recovering $\Phi(\beta)$ exactly. To see this, we begin by defining
\begin{equation}
    R(E;\beta) = (\beta-\beta_{\rm sp}(E))E - \ln \Phi(\beta_{\rm sp}(E))
\end{equation}
such that (making use of Eq.~\ref{eq:Energy_Definition}) we have
\begin{equation} \label{eq:Rprime}
    R'(E;\beta) = \beta - \beta_{\rm sp}(E)
\end{equation}
and 
\begin{equation}
    R''(E;\beta) =  - \beta'_{\rm sp}(E).
\end{equation}
Setting Eq.~\ref{eq:Rprime} to zero, we see that the steepest-descent condition is
\begin{equation}
    \beta=\beta_{\rm sp}(E_{\rm sp}) ,
\end{equation}
i.e.~$E_{\rm sp}=\mathcal{E}(\beta)$
and that
\begin{equation}
    R''(E_{\rm sp};\beta) =  - \beta'_{\rm sp}(E_{\rm sp}) = - \left(\frac{\mathrm{d}\mathcal{\beta_{\rm sp}}}{\mathrm{d}E}\right)_{E=E_{\rm sp}} = - \left(\frac{\mathrm{d}\mathcal{E}}{\mathrm{d}\beta}\right)^{-1}.
\end{equation}
Combining these results, we can perform the Laplace transform, Eq.~\ref{eq:LT}, asymptotically  
\begin{equation}
   \frac{1}{2\pi\lambda} {\int_{-\infty}^\infty}  \sqrt{2\pi\lambda}\,\Bigg(\!-\!\frac{\mathrm{d}\mathcal{E}}{\mathrm{d}\beta}\Bigg)^{-\nicefrac{1}{2}}_{\beta=\beta_{\rm sp}} \, e^{-R(E;\beta)/\lambda} \, \mathrm{d} E
\sim \frac{1}{2\pi\lambda} \sqrt{2\pi\lambda} \Bigg(\!-\!\frac{\mathrm{d}\mathcal{E}}{\mathrm{d}\beta}\Bigg)^{-\nicefrac{1}{2}} \sqrt{2\pi\lambda}\Bigg(\!-\!\frac{\mathrm{d}\mathcal{E}}{\mathrm{d}\beta}\Bigg)^{\nicefrac{1}{2}} e^{\ln \Phi(\beta) /\lambda} ,
\end{equation}
which upon setting $\lambda=1$, recovers $\Phi(\beta)$.

Now we will make use of this consistency to derive an approximation to the Laplace transform with a modified lower bound,
\begin{equation}
    \Phi(\beta;E_{\rm min}) = \frac{1}{2\pi\hbar} \int_{E_{\rm min}}^\infty N(E) \, e^{-\beta E} \,\mathrm{d} E.
\end{equation}
To achieve this, we will insert the expression for the stationary-phase approximation to the inverse Laplace transform  (and reintroduce the perturbative parameter $\lambda$)
\begin{equation}
    \Phi(\beta;E_{\rm min}) \sim \frac{1}{\sqrt{2\pi\lambda}} \int_{E_{\rm min}}^\infty  \,\Bigg(\!-\!\frac{\mathrm{d}\mathcal{E}}{\mathrm{d}\beta}\Bigg)^{-\nicefrac{1}{2}}_{\beta=\beta_{\rm sp}} \, e^{-R(E;\beta)/\lambda}  \, \mathrm{d} E
\end{equation}
and then integrate asymptotically via steepest descent, but using Bleistein's method to account for the lower limit. A review of Bleistein's method is given in the following section; here we simply make use of the result, which for the present integral is
\begin{equation}
\begin{aligned}
    \Phi(\beta;E_{\rm min}) \sim \frac{1}{\sqrt{2\pi\lambda}} &\Bigg[   \,\Bigg(\!-\!\frac{\mathrm{d}\mathcal{E}}{\mathrm{d}\beta}\Bigg)^{-\nicefrac{1}{2}} \, \sqrt{\frac{2\pi\lambda}{R''(\mathcal{E};\beta)}} e^{-R(\mathcal{E};\beta)/\lambda} \frac{1}{2} \mathrm{erfc}\left(\mathrm{sgn}(E_{\rm min}-\mathcal{E})\sqrt{\frac{\Delta R}{\lambda}}\right)\\
    &+\lambda e^{-R(E_{\rm min};\beta)/\lambda} \frac{\mathrm{sgn}(\mathcal{E}-E_{\rm min})}{\sqrt{2R''(\mathcal{E};\beta)\Delta R}}\Bigg(\!-\!\frac{\mathrm{d}\mathcal{E}}{\mathrm{d}\beta}\Bigg)^{-\nicefrac{1}{2}} + \lambda e^{-R(E_{\rm min};\beta)/\lambda} \frac{1}{R'(E_{\rm min};\beta)}\Bigg(\!-\!\frac{\mathrm{d}\mathcal{E}}{\mathrm{d}\beta}\Bigg)^{-\nicefrac{1}{2}}_{\beta=\beta_{\rm sp}(E_{\rm min})} \Bigg].  
    \end{aligned}
\end{equation}
This can then be simplified by defining
\begin{equation}
    \beta_{\rm min} = \beta_{\rm sp}(E_{\rm min})
\end{equation}
and
\begin{equation}
    \Delta R = R(E_{\rm min};\beta)-R(\mathcal{E};\beta) = (\beta - \beta_{\rm min})E_{\rm min} - \ln \Phi(\beta_{\rm min}) + \ln \Phi(\beta)
\end{equation}
along with setting $\lambda=1$ to give 
\begin{equation}
\begin{aligned}
    \Phi(\beta;E_{\rm min}) &\approx \Phi(\beta) \frac{1}{2} \mathrm{erfc}\left(\mathrm{sgn}(E_{\rm min}-\mathcal{E})\sqrt{\Delta R}\right)\\ &+ \frac{1}{\sqrt{2\pi}}\Phi(\beta_{\rm min})\,e^{-(\beta-\beta_{\rm min})E_{\rm min}} \left(\frac{\mathrm{sgn}(\mathcal{E}-E_{\rm min})}{\sqrt{2\Delta R}}+\frac{1}{\beta-\beta_{\rm min}} \Bigg(\!-\!\frac{\mathrm{d}\mathcal{E}}{\mathrm{d}\beta}\Bigg)^{-\nicefrac{1}{2}}_{\beta=\beta_{\rm min}}\right) \label{Eq:final_1}
    \end{aligned}
\end{equation}
or equivalently
\begin{equation}  \label{Eq:final_2}
\begin{aligned}
    \Phi(\beta;E_{\rm min}) &\approx \Phi(\beta) - \Phi(\beta)\frac{1}{2}  \mathrm{erfc}\left(\mathrm{sgn}(\beta_{\rm min}-\beta)\sqrt{\Delta R}\right)\\ &+ \frac{1}{\sqrt{2\pi}}\Phi(\beta_{\rm min})\,e^{-(\beta-\beta_{\rm min})E_{\rm min}} \left(\frac{\mathrm{sgn}(\beta_{\rm min}-\beta)}{\sqrt{2\Delta R}}+\frac{1}{\beta-\beta_{\rm min}} \Bigg(\!-\!\frac{\mathrm{d}\mathcal{E}}{\mathrm{d}\beta}\Bigg)^{-\nicefrac{1}{2}}_{\beta=\beta_{\rm min}}\right).
    \end{aligned}
\end{equation}
Dividing Eq.~\ref{Eq:final_2} by $Z_r(\beta)$, one recovers the expression given in the main text for $k^{\rm SLTA}(\beta;E_{\rm min})$ [Eq.~13].

Note that although we have derived this result by first performing the inverse Laplace transform and then performing the Laplace transform by steepest descent with a modified lower bound, it would also be possible to derive the same expression by analysis of the corresponding convolution integral.

\section{Summary of Bleistein's Method}

For an integral of the form
\begin{equation}
    I(\lambda) = \int_{x_{\rm min}}^\infty g(x) \, e^{-f(x)/\lambda} \, \mathrm{d}x ,
\end{equation}
where the stationary point, $x^\star$, of $f(x)$ may lie either inside or outside of the integration range, Bleistein's method allows one to calculate a uniform asymptotic expansion that is continuously valid as the stationary point moves through the boundary.
Following Bleistein's approach,\cite{Bleistein1966Uniform,RWong1989UniformAsymptotics} one begins by making the substitution
\begin{equation}
    f(x) = \tfrac{1}{2} u^2 -bu + c
\end{equation}
where
\begin{align}
    b &= %
    \mathrm{sgn}(x^\star-x_{\rm min})\sqrt{2\Delta f}
    \\
    \Delta f &= f(x_{\rm min}) - f(x^\star)
    \\
    c &= f(x_{\rm min}) ,
\end{align}
such that $u=0$ corresponds to $x=x_{\rm min}$ and $u=b$ to $x=x^\star$. This allows us to express the integral in the form
\begin{equation}
   I(\lambda) =  \int_0^\infty h(u) \, e^{-(u^2/2-bu+c)/\lambda} \, \mathrm{d}u ,
\end{equation}
where
\begin{equation}
    h(u) = g\big(x(u)\big) \frac{\mathrm{d}x}{\mathrm{d}u}.
\end{equation}
Rewriting $h(u)=h(b)+(u-b)\frac{h(b)-h(0)}{b} + r(u)$, the remainder term, $r(u)$, can be ignored at leading order and the integral can be performed analytically to give 
\begin{equation}
    I(\lambda) \sim e^{-c/\lambda} h(b) \sqrt{2\pi\lambda} \, e^{b^2/2\lambda} \frac{1}{2}\mathrm{erfc}\!\left(\frac{-b}{\sqrt{2\lambda}}\right) + \lambda\frac{h(b)-h(0)}{b} e^{-c/\lambda}.
\end{equation}
This can be recast in terms of $x$ by noting that
\begin{equation}
    \frac{\mathrm{d}u}{\mathrm{d}x} = \mathrm{sgn}(x-x^\star) \frac{f'(x)}{\sqrt{2f(x)-2f(x^\star)}} ,
\end{equation}
such that (by carefully taking limits)
\begin{equation}
    h(b) = \frac{g\big(x^\star\big)}{\sqrt{f''(x^\star)}}
\end{equation}
and
\begin{equation}
    h(0) = g(x_{\rm min}) \mathrm{sgn}(x_{\rm min}-x^\star) \frac{\sqrt{2f(x_{\rm min})-2f(x^\star)}}{f'(x_{\rm min})} = -\frac{g(x_{\rm min}) b}{f'(x_{\rm min})}.
\end{equation}
Finally, noting that $-b^2/2+c=f(x^\star)$, we obtain the final result, which is the first term in the uniform asymptotic expansion valid for any value of $x^\star-x_{\rm min}$:
\begin{equation}
\begin{aligned}
   I(\lambda) &\sim g(x^\star) \sqrt{\frac{2\pi\lambda}{f''(x^\star)}} e^{-f(x^\star)/\lambda} \, \frac{1}{2}\mathrm{erfc}\left(\mathrm{sgn}(x_{\rm min}-x^\star)\sqrt{\frac{\Delta f}{\lambda}}\right) \\ & + \lambda e^{-f(x_{\rm min})/\lambda} \frac{\mathrm{sgn}(x^\star-x_{\rm min})g(x^\star)}{\sqrt{2 f''(x^\star) \Delta f}} +\lambda e^{-f(x_{\rm min})/\lambda}\frac{g(x_{\rm min})}{f'(x_{\rm min})}.
   \end{aligned}
\end{equation}

\bibliographystyle{achemso}
\bibliography{references,extra_refs} %